\documentclass[preprint]{aastex}

\usepackage{amssymb,amsmath}
\usepackage{booktabs}

\begin{document}
\slugcomment{March 10, 2016: Accepted to ApJ}

\title{Highly Excited H$_2$ in Herbig-Haro 7:\\Formation Pumping in Shocked Molecular Gas?}

\author{R. E. Pike\altaffilmark{1}}
\author{T. R. Geballe\altaffilmark{2}}
\author{M. G. Burton\altaffilmark{3,4}}
\author{A. Chrysostomou\altaffilmark{5}}

\altaffiltext{1}{Department of Physics and Astronomy, University of Victoria, Victoria, BC, Canada}
\altaffiltext{2}{Gemini Observatory, Hilo, HI, USA}
\altaffiltext{3}{School of Physics, University of New South Wales, Sydney, Australia}
\altaffiltext{4}{Armagh Observatory and Planetarium, College Hill, Armagh BT61 9DG, Northern Ireland, UK}
\altaffiltext{5}{Centre for Astrophysics Research, University of Hertfordshire, UK}

\begin{abstract}

We have obtained $K$-band spectra at R$\sim$5,000 and angular resolution  0$\farcs$3 of a section of the Herbig-Haro 7 (HH7) bow shock, using the Near-Infrared Integral Field Spectrograph at Gemini North. Present in the portion of the data cube corresponding to the brightest part of the bow shock are emission lines of H$_2$ with upper state energies ranging from $\sim$6,000 K up to the dissociation energy of H$_2$, $\sim$50,000 K.  Because of low signal-to-noise ratios, the highest excitation lines cannot be easily seen elsewhere in the observed region. However, excitation temperatures, measured  throughout much of the observed region using lines from levels as high as 25,000~K, are a strong function of upper level energy, indicating that the very highest levels are populated throughout. The level populations in the brightest region are well fit by a two-temperature model, with 98.5\%\ of the emitting gas at T=1800~K and 1.5\%\ at T=5200~K. The bulk of the H$_2$ line emission in HH7, from the 1,800~K gas, has previously been well modeled by a continuous shock, but the 5,200~K component is inconsistent with standalone standard continuous shock models.  We discuss various possible origins for the hot component and suggest that this component is H$_2$ newly reformed on dust grains and then ejected from them, presumably following dissociation of some of the H$_2$ by the shock. 
 
\end{abstract}

\keywords{ISM: lines and bands -- ISM: molecules -- line: identification --  molecular processes -- shock waves}

\section{Introduction}

Shock waves in molecular clouds are often created by the collisions of supersonic outflowing gas from pre-main sequence stars with ambient molecular gas from the clouds out of which the stars formed. The shock structure and physics are commonly studied using the emission lines of molecular hydrogen (H$_2$), which is dissociated when collisions of H$_2$ and atoms or molecules occur at speeds exceeding 20-24~km~s$^-$$^1$ \citep{hol77,kwa77,lon77}. In many cases the relative velocities of the outflow and cloud far exceed this limit \citep[e.g.,][] {nad79}, yet in many shocked clouds, strong H$_2$ lines with broad velocity profiles are seen. 

The commonly accepted solution to this puzzle has been to invoke continuous shocks (C-shocks), in which the ambient cloud is heated and gradually accelerated, first by collisions with ions in a magnetic precursor,  and then by previously swept up molecular gas \citep{dra80,dra82,che82}. The slower acceleration can result in a low peak temperature (typically 2,000~K) of the shock-heated gas and the rough maintenance of that temperature over a long column of shocked gas \citep{smi91}, reducing the amount of dissociation. Shocks speeds can be as high as  $\sim$45 km~s$^{-1}$ before complete dissociation of the molecular gas occurs \citep{dra83}.The expected relative line intensities in such low temperature C-shocks are in contrast to jump shocks (J-shocks), in which for velocities exceeding several km~s$^{-1}$ the molecular gas is suddenly heated to much higher temperatures and then quickly cools, resulting in a wide range of temperatures in the post-shock gas.  Excitation of H$_2$ by ultraviolet radiation produced in the shock is another means by which the H$_2$  can be excited into emission in its ro-vibrational lines \citep{bla87,ste89,bur90}. UV excitation produces a spectral signature distinct from that of collisionally heated H$_2$. An additional possibility is dissociation of the H$_2$ at the shock front, followed by reformation of H$_2$ on dust grains behind the shock front, with radiative cascading of H$_2$ ejected from the grains down the vibrational ladder, eventually resulting in vibrational local thermodynamic equilibrium (LTE), as well as collisional thermalization of rotational levels \citep{flo03}. 

Early attempts to distinguish between various shock types and cooling zones, which utilized measured relative intensities of H$_2$ lines mostly in the Orion Molecular Cloud \citep{bec78} and comparisons of velocity profiles at Peak 1 \citep{moo90}, all found that that by themselves C-shock models were incompatible with the observations \citep{bra88,bur89a,bra89}. Probably the most significant of these tests \citep{bra88} was the measurement of 19 ro-vibrational and pure rotational lines between 2~$\mu$m and 4~$\mu$m at Peak 1, whose upper energy levels ranged up to $\sim$25,000~K. The extinction-corrected relative intensities appeared to be more consistent with post-shock cooling starting at very high temperatures, i.e., in the absence of a shock-softener.  

All of the tests mentioned above were made with angular resolutions of several arc-seconds, so it is possible that each observation took in a large number of C-shocks producing widely different peak temperatures, which might have blurred a potentially observable distinction between C-shocks and other possible explanations. More recent observations \citep[e.g.][]{fer95} used smaller apertures, but covered only small regions of the shocked gas. In addition, the lower instrumental sensitivities at those earlier times, although allowing the detection of a number of weak lines from energy levels much higher than expected in C-shocked gas, may have prevented a more complete understanding of the line emission. Now, more than two decades later, larger telescopes and improvements in instrumentation have made it possible to observe H$_2$ lines at an order of magnitude higher angular resolution and with much higher sensitivity, and thus to begin to examine shocked interstellar molecular gas in greater detail.  

\begin{figure}[h!]
\begin{center}
\includegraphics[width=1\textwidth]{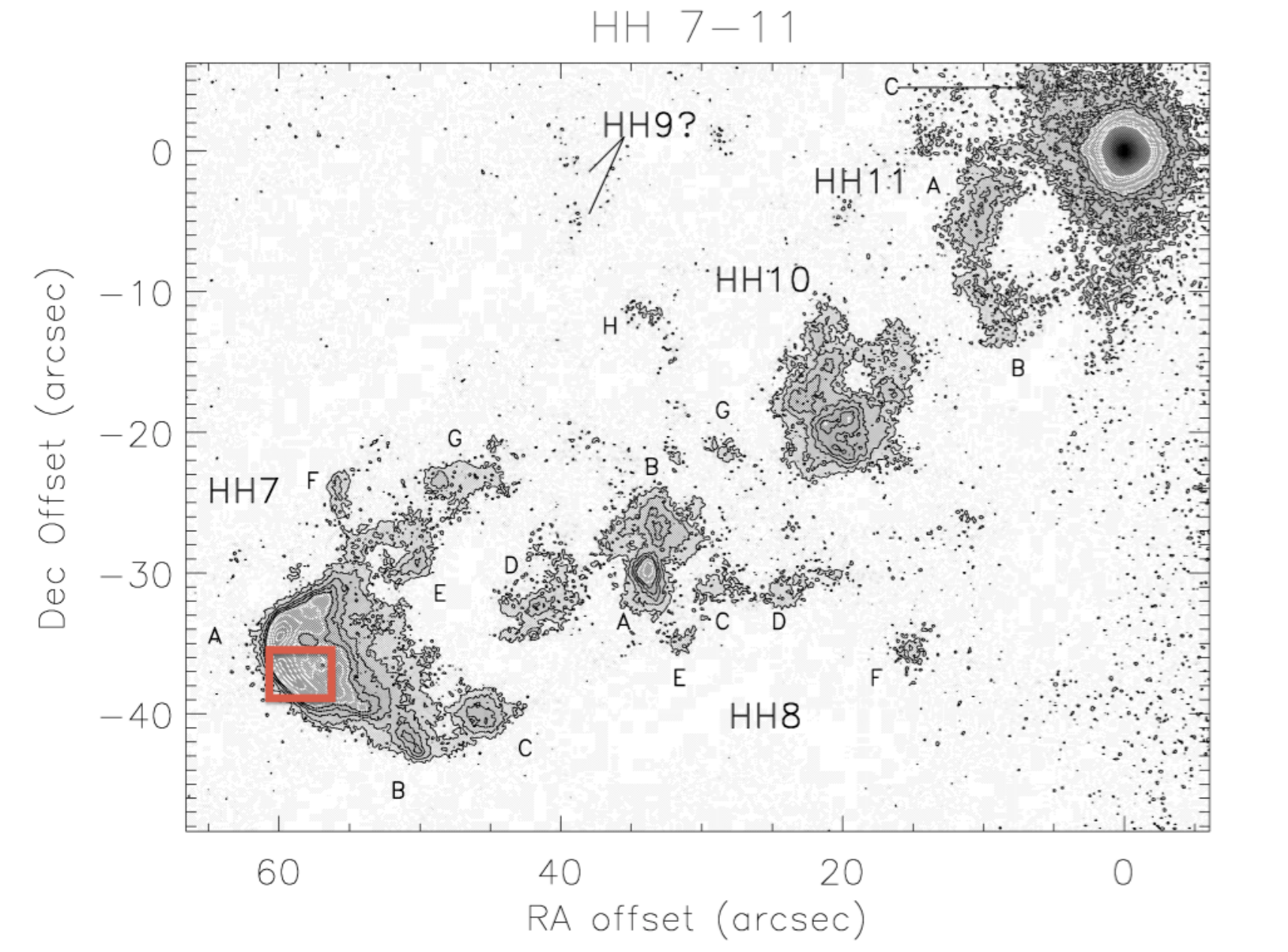}
\caption{Contour map and image of the H$_2$ 1--0 $S$(1) line intensity in HH 7-11, from \citet{kha03}. The red rectangle at lower left denotes the location of the NIFS data cube. The bright star at upper right is SVS~13, the probable source of the outflow. }
\label{map}
\end{center}
\end{figure}

The collimated outflow (jet) from the protostar SVS13 has produced a collection of Herbig-Haro objects, HH7--HH11, with a classic bow shock, HH7, at its eastern end, as shown in  Fig.~\ref{map}. The bow shock is bright in line emission from molecular hydrogen \citep{gar90}. As in the shock in the Orion Molecular Cloud, H$_2$ lines from upper energy levels as high as $\sim$25,000~K have been detected in HH7 \citep{fer95}. At a distance of only 250~pc \citep{eno06}, HH7 is at about half the distance of the Orion Molecular Cloud and its morphology projected onto the sky is much simpler. Estimates of the inclination of HH7 to the line of sight vary considerably, but suggest an angle of $\sim$40~$\deg$ \citep{bac90,car93,fer95,kha03,smi03}.  The proximity of HH7, the relative simplicity of the geometry of the shock, and the brightness of the line emission suggested to us that HH7 would be a favorable object for investigating the detailed physics in a molecular shock. 

\section{Observations and Data Reduction}

Integral field spectra of a small section of the HH7 bow shock (shown in Fig.~\ref{map}), adjacent to the cap of the bow shock, were acquired at Frederick C. Gillett Gemini North Telescope on Maunakea, Hawaii using the facility Near-Infrared Field Spectrometer (NIFS) \citep{mcg02} on UT 16 October 2007 and 22 December 2007, for program GN-2007B-Q-47. The field of view of NIFS is 3$\farcs$0 $\times$3$\farcs$0 with individual spaxel sizes of 0$\farcs$04 $\times$ 0$\farcs$10. The grating in NIFS, which provides a spectral resolving power of 5000, corresponding to a velocity resolution of 60~km~s$^{-1}$, and a velocity sampling of 30~km~s$^{-1}$, was positioned to cover the wavelength range 2.01-2.45~$\mu$m. Adaptive optics was not employed, as no nearby star of sufficient optical brightness is available for natural guide star adaptive optics and no suitable tip-tilt star is available near HH7 to allow laser guide star adaptive optics.  The observations were made in photometric skies with seeing of 0$\farcs$35 or better (FWHM) in the $K$ band.  The precipitable water vapor above Maunakea was $\sim$3~mm on the first night and $\sim$2~mm on the second. 

The observations of HH7 consist of 10 exposures of 552 seconds each on 16 October 2007 and 16 exposures of 660 seconds each on 22 December 2007, with half of the exposures on target and half on adjacent sky. The target position was acquired each night by offsetting from a nearby star.  A telluric standard (HIP 10559; A0V) was observed immediately prior to HH7 on each night.  

Baseline reductions of the NIFS data cubes were performed using Gemini software within the NIFS IRAF package. The calibration data included $K$ band flats, spectra of argon and xenon arc lamps, darks, and flats observed through a Ronchi mask. The science and telluric standard data were both dark-subtracted and divided by the flat field. The wavelength solutions to the arc spectra were applied and the data shifted to a heliocentric wavelength scale.  Some bad pixels were replaced by  interpolated values. The spectra of HH7 were divided by the spectrum of the reference star to remove the effects of atmospheric absorption and instrumental transmission and to provide flux calibration.

In the $K$ band the bow shock is visible only in H$_2$ emission, which has no point-like distinguishing characteristics. Consequently, blind offsetting from a nearby visible point source was used to position the field of view of NIFS. This resulted in a small positional difference between the two nights. During each night positional offsets were minimal. After inspection by eye the images from one night were shifted to align the two data sets and then all images were combined. The alignment is believed to be accurate to better than a spaxel. This cube was then trimmed to remove outer regions of low signal-to-noise ratios, mostly regions where the data cubes from the two nights did not overlap. The final data cube has  0$\farcs$05~$\times$~0$\farcs$05~ elements covering a 3$\farcs$6~$\times$~2$\farcs$9 region (RA~$\times$~Dec) and spans the wavelength interval 2.011-2.446~$\mu$m. The spectra in the cube have been multiplied by a blackbody function having the temperature of the A0V reference star (9480~K) to produce final ratioed spectra in dimensions of F$_\lambda$. Approximate flux calibration was obtained by scaling the NIFS spectra at the peak of the H$_2$ line emission to the observations by \citet{smi03}. 

\begin{figure}[h!]
\begin{center}
\includegraphics[width=1\textwidth]{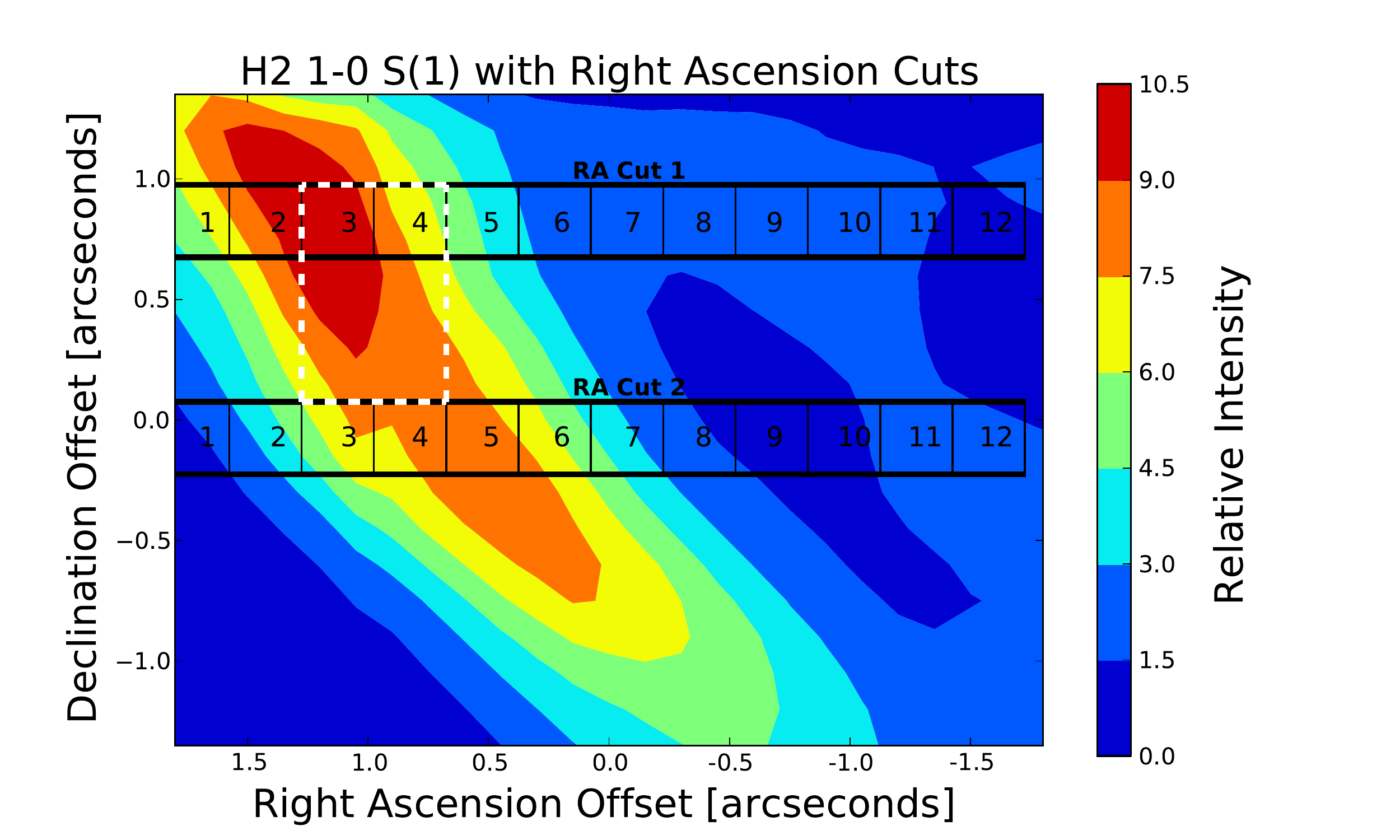}
\caption{Intensity of the H$_2$ 1--0 $S$(1) line in the observed region. North is up and East is to the left. The origin of the plot corresponds to RA = 3:29:08.39, Dec = +31:15:27:45 (J2000) with an estimated uncertainty of 0$\farcs$25. The sets of 0\farcs30~$\times$~0\farcs30 boxes are regions where spectra and line profiles are shown in Figs. \ref{lines1} and \ref{lines2}.  The white dashed box shows the location of the 0\farcs6$\times$0\farcs9 area whose spectrum is shown in Fig.~\ref{binned_spec}.}
\label{S1line_map}
\end{center}
\end{figure}

\begin{figure}[h!]
\begin{center}
\includegraphics[width=1.1\textwidth]{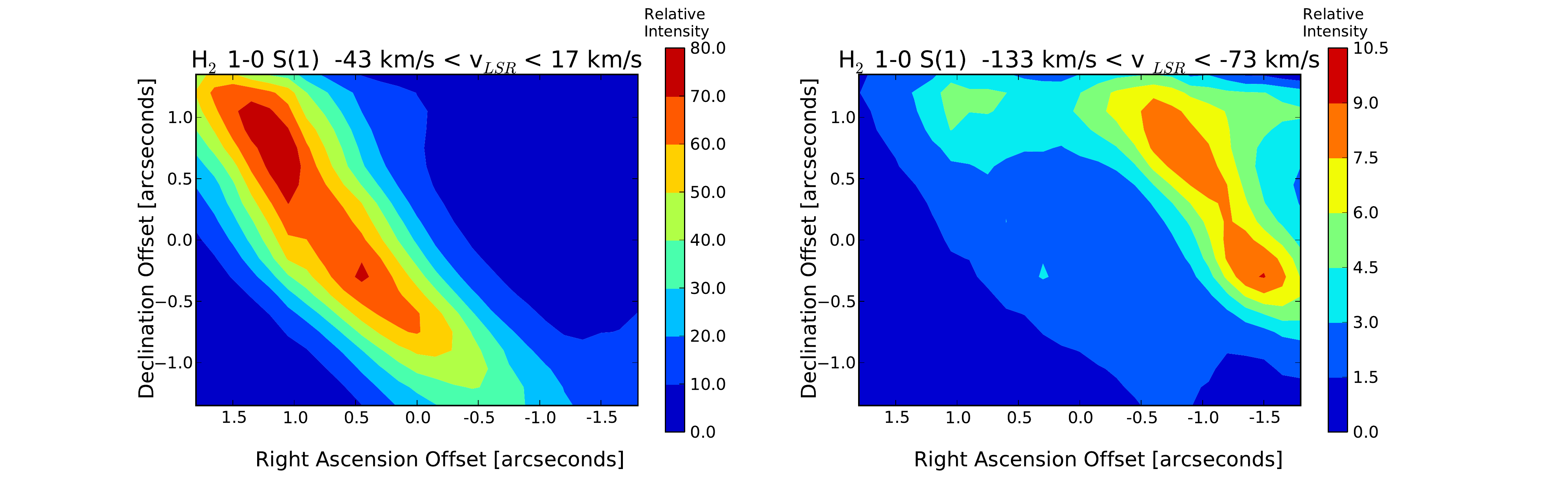}
\caption{Intensity maps of the 1--0 $S$(1) line in two velocity intervals centered at -13 km~s$^{-1}$ and -103 km~s$^{-1}$. Note the difference in intensities between the two maps.}
\label{velocity}
\end{center}
\end{figure}

\begin{figure}[h!]
\begin{center}
\includegraphics[width=1.18\textwidth]{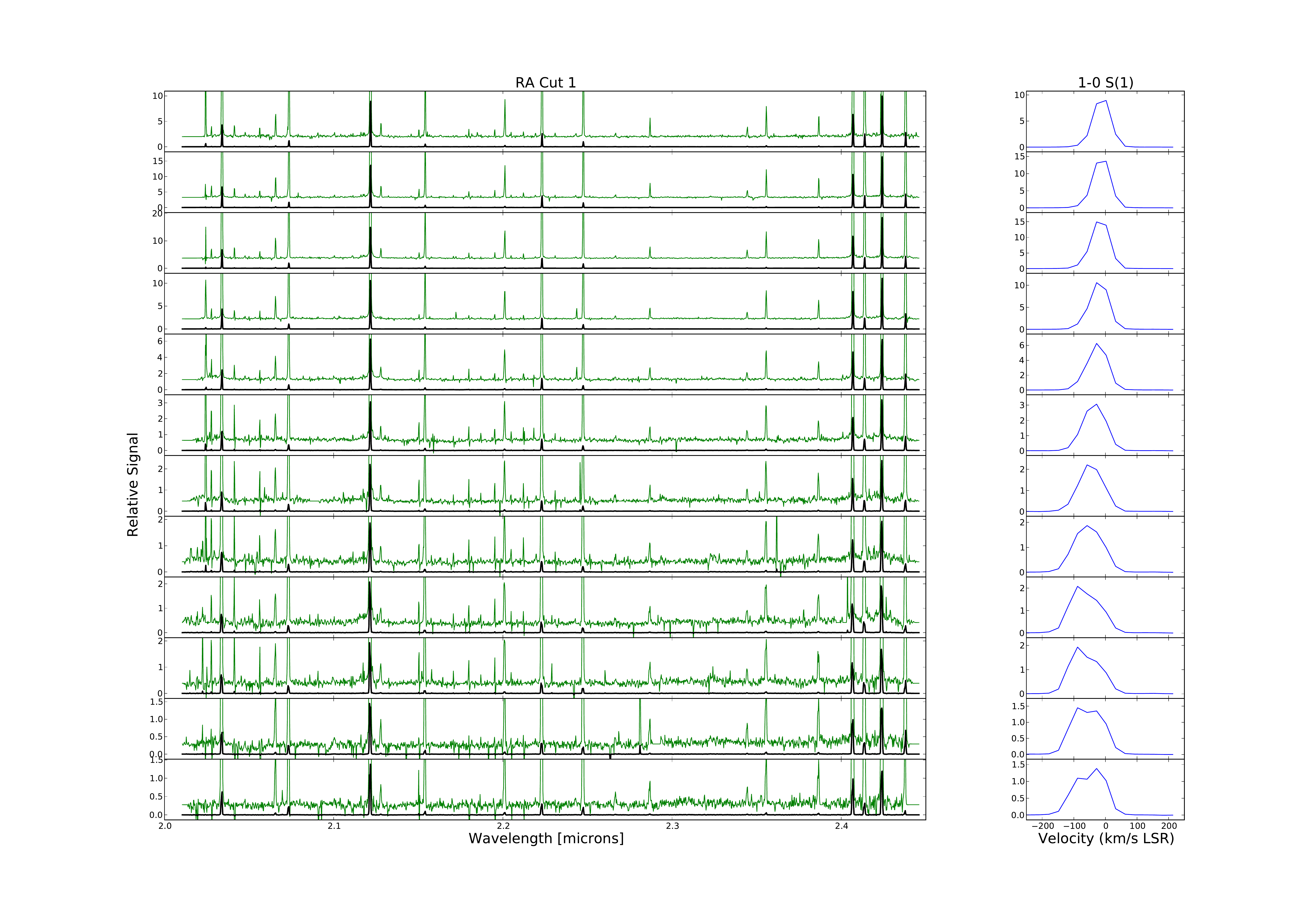}
\caption{Spectra and line profiles from the upper line cut in Fig. \ref{S1line_map}; regions from left to right in Fig. \ref{S1line_map} are shown from top to bottom here.  Left: 2.01-2.45~$\mu$m spectra (black lines).  Green lines are spectra magnified by a factor of 150 and offset vertically to show weaker lines. Right: velocity profiles of the 1--0 $S$(1) line.}
\label{lines1}
\end{center}
\end{figure}

\begin{figure}[h!]
\begin{center}
\includegraphics[width=1.18\textwidth]{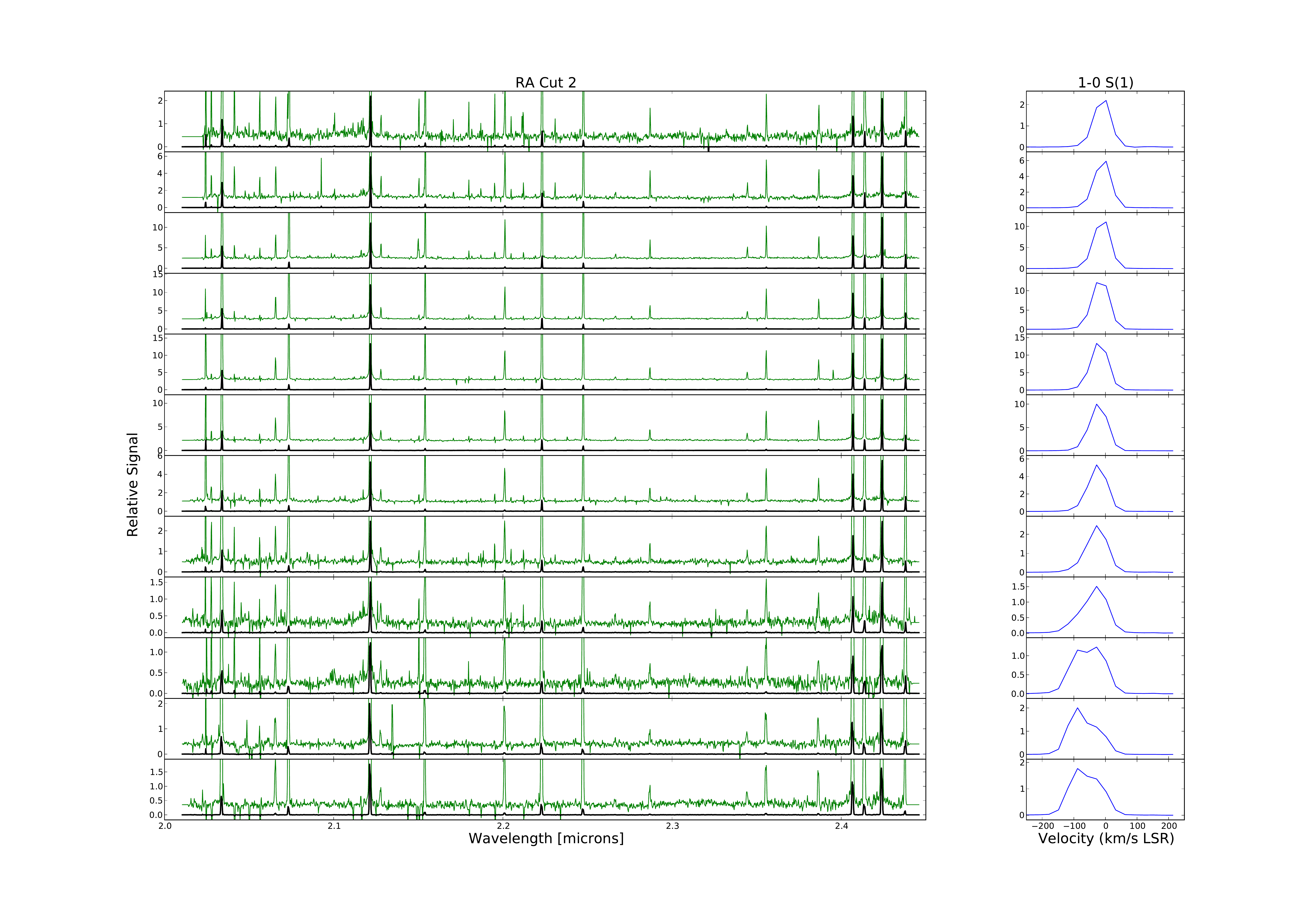}
\caption{Spectra and line profiles from the lower line cut in Figure \ref{S1line_map}. Other details as per Fig.~\ref{lines1}.}
\label{lines2}
\end{center}
\end{figure}

\section{Overview of the Data}

Figure \ref{S1line_map} is a map of the 1--0 $S$(1) line intensity (integrated over its full velocity profile) in the observed region, and Fig.~\ref{velocity} contains separate intensity maps of the two velocity components that make up the bulk of the line emission. The morphology of the total line emission closely matches that of  Fig.~\ref{map} \citep{kha03}, with the leading edge of the bow on the eastern edge of the field of view and the brightest emission approximately 0.8 arc-seconds to the west of that edge. Further to the west a second and considerably fainter ridge of line emission is present across the region.  At the location of the tip of the bow (top left of Fig. \ref{S1line_map}) the intensity is somewhat lower than along the side of the bow, also consistent with the image of \citet{kha03}. 

Figures \ref{lines1} and \ref{lines2} show sets of spectra of 0\farcs30~$\times$~0\farcs30 regions at  fixed declinations, crossing the bow shock in right ascension (with decreasing RA from top to bottom in these figures). As discussed below, all of the observed emission lines are from H$_2$.  Velocity profiles of the 1--0 $S$(1) line in each region are shown on the right sides of the figures.  The locations of these spectra are marked in Fig.~\ref{S1line_map}. The spectra reveal that the peak intensities of all of the stronger lines vary roughly in lockstep as the individual line intensities change by over an order of magnitude across the region. The 1--0 $S$(1) line profiles show that the line emission occurs largely in two velocity intervals. Intensity maps of these two components are shown in Fig.~\ref{velocity}. The brightest emission, in the eastern part of the data cube, peaks near $-$15~km~s$^{-1}$ (all velocities are in LSR, the Local Standard of Rest). The line profiles there are roughly symmetric as viewed at the instrumental resolution. Behind the front (to the west) this velocity component weakens rapidly to one-tenth its peak intensity, the line broadens, and the line centroid shifts to more negative velocities, as a second component centered near $-100$~km~s$^{-1}$ appears.  Due to the intrinsic velocity widths of these components and the modest resolution of NIFS the two components are only partially resolved. The blue-shifted second velocity component appears in the total intensity map in  Fig.~\ref{S1line_map} as the aforementioned ridge to the west of the brightest region, whereas in Fig.~\ref{velocity} it is clearly seen as a separate morphological feature. The location of this second component is consistent with the findings of \citet{dav00}, who observed HH7 at higher spectral resolution but lower angular resolution.  Both the location and the high negative velocity suggests that this component is associated with the reverse shock in the jet, created as the outflow from SVS13 begins to impact the bow shock. 

Close inspection of the profiles of the strongest lines reveal very weak wings extending $\pm$300~km~s$^{-1}$ from line centers. However, spectra of strong arc lamp lines show similar wings, so the wings are instrumental artifacts. 

\begin{figure}[h!]
\centering
\begin{center}
\includegraphics[width=1.18\textwidth]{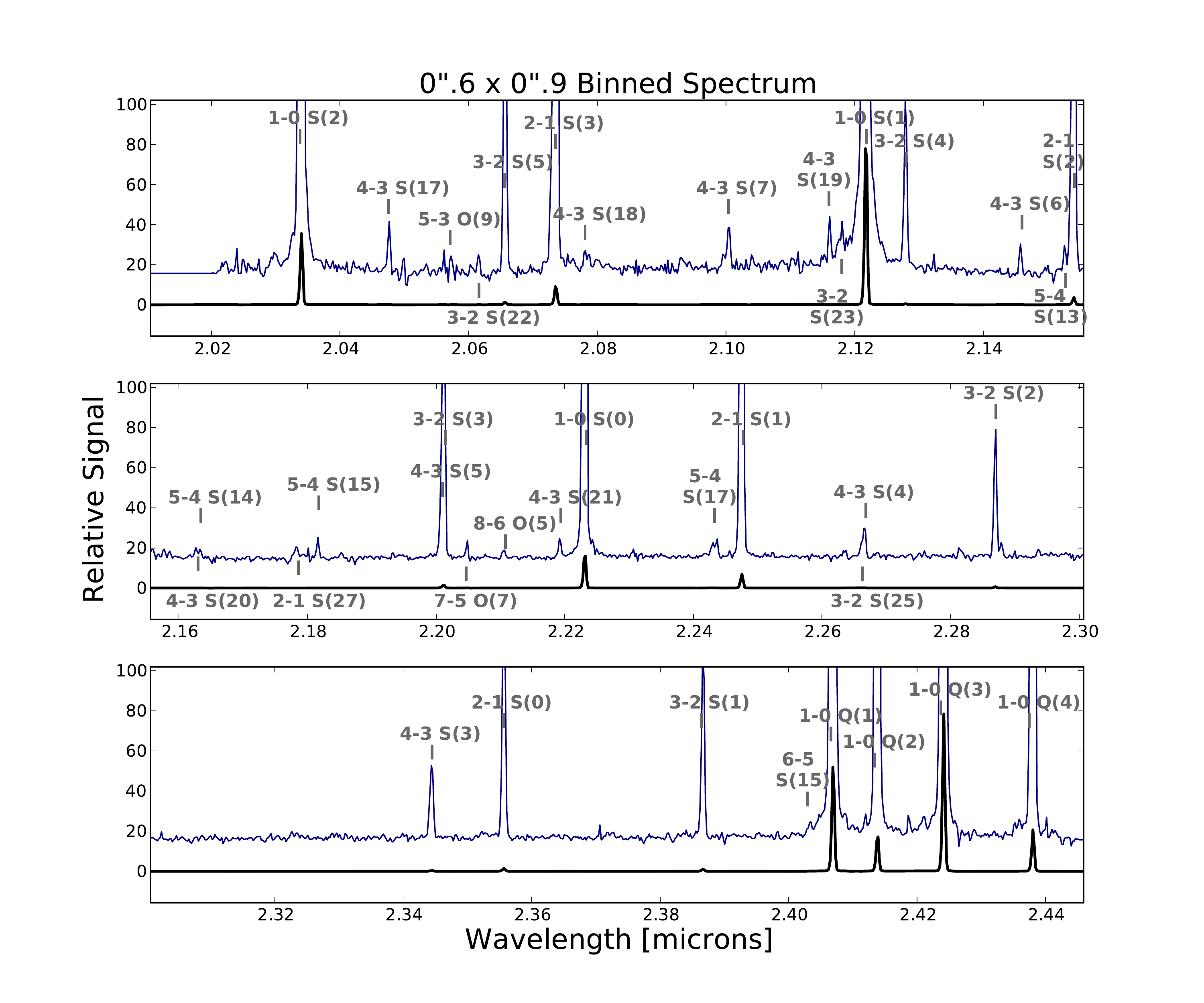}
\caption{Spectrum of 0\farcs6~$\times$~0\farcs9 region delineated by white dashed lines in Figure \ref{S1line_map}.  The blue trace is the spectrum magnified by a factor of 150 and offset vertically to show weaker lines. H$_2$ lines are labelled; tick marks correspond to laboratory wavelengths.}
\label{binned_spec}
\end{center}
\end{figure}

Thirty-eight emission lines due to H$_2$ are detected in the data cube. Most of the lines can be seen in the spectra contained in Figs. \ref{lines1} and \ref{lines2}, but in order to provide higher signal-to-noise ratios on the very weak lines for more detailed analysis, a spectrum of the 0\farcs6~$\times$~0\farcs9 region centered at RA and Dec offsets of -1\farcs0 and 0\farcs5, respectively was extracted and is shown in a more expanded form in Fig.~\ref{binned_spec}.  The lines and their relevant parameters are listed in Table~\ref{lines}. The dynamic range, between the strongest (1--0 $S$(1)) and weakest detected lines, is approximately 1,500. It is this high dynamic range together with the high sensitivity of NIFS that makes possible the new analysis presented in this paper.

The upper state energy levels of the observed lines range from 6471 K for the 1--0 $S$(0) line to 52,643~K for the 2--1 $S$(27) line. For comparison the dissociation energy of H$_2$ from its ground state \citep{her70, bal92}, has an equivalent temperature of 51,967~K, apparently exceeded by that of the observed $v$=2 $J$=29 level. No spontaneous transition rate has been published for the 2--1 $S$(27)  line. 

The detection of roughly a dozen infrared lines from ro-vibrational states that are 40,000-50,000~K above the ground state is surprising. Previously in regions of shocked molecular hydrogen, the highest excitation energies detected were  $\sim$25,000~K. Most of the highest excitation lines are quite weak, but both the excellent matches of their observed wavelengths to their predicted wavelengths and their morphologies, which are similar to those of the stronger lines in the brightest regions, argue conclusively for our identifications. In the spectra in Fig.~\ref{lines1} and Fig.~\ref{lines2} they are only detected in the region of strongest line emission. Some other very weak features are evident in Fig.~\ref{binned_spec}, but because these do not exhibit the same morphology, having a patchy spatial distribution at best, we interpret them to be due to noise fluctuations or contamination by cosmic rays.

All but three of the H$_2$ lines in Fig.~\ref{binned_spec} are fundamental ($\Delta$$v$=1) band transitions. The other three are first overtone transitions. Three of the emission features, at 2.163, 2.201, and 2.267~$\mu$m, are blends of pairs of lines.  In the case of the strongest blended feature, the 4--3 $S$(5) line at 2.2010~$\mu$m and the 3--2 $S$(3) line at 2.2014~$\mu$m, both deconvolution and modeling (see Section 4.2) show that the higher excitation line contributes approximately 25\% of the signal. 

While many of the H$_2$ lines are clear of interference from moderate strength or strong telluric absorption lines, some are affected, as indicated in Table~\ref{lines}. For medium strength telluric lines we estimate that calibrated line strengths are affected, at most, at the few percent level by uncertainty in the correction; in the second instance the effect could be much larger, and our analysis has avoided such lines.  In addition, residual OH sky lines, present in the data cube due to the sky emission varying during the observations may seriously contaminate a few of the H$_2$ lines, as indicated in the table; these lines also are avoided in our analysis.  Uncertainty in the extinction correction (discussed in Section 4.1) also contributes to uncertainty in the values in column 6 of Table~\ref{lines}, but its effect on the scientific thrust of this paper is negligible.

\begin{table}[http]
\caption{Observed H$_2$ Emission Lines $^a$}
\label{lines}
\begin{center}
\tiny
\begin{tabular}{|c|c|c|c|c|c|c|}
\hline
Line & Rest $\lambda$  & Upper Level & A & Observed & Dereddened & Notes \\
ID & vac. $\mu$m$^b$ & Energy (K)$^b$ & 10$^{-7}$s$^{-1}$ $^c$ & Intensity$^d$ & Rel. Intensity$^e$ & \\
\hline
1--0 $S$(2)   & 2.0338 & 7,585 & 3.98 & 80.8 & 43.0 & o \\
4--3 $S$(17) & 2.0475 & 42,022 & 13.4 &  0.39 & 0.20 & s \\
5--3 $O$(9)   & 2.0571 & 30,064 & 1.56 & 0.13 & 0.07 & s \\
3--2 $S$(22) & 2.0616 & 47,012 & - & 0.13 & 0.07 & s \\
3--2 $ $S(5)   & 2.0656 & 20,857 & 4.50 & 3.2 & 1.7 & s \\
2--1 $S$(3)   & 2.0735 & 13,890 & 5.77 & 23.4 & 12.0 & o,s \\ 
4--3 $S$(18) & 2.0781 & 43,614 & 16.7 & 0.17 & 0.09 & m \\
4--3 $S$(7)   & 2.1004 & 27,707 &  2.98 & 0.34 & 0.17 & \\
4--3 $S$(19) & 2.1163 & 45,204 & 19.9 & 0.35 & 0.17 &  \\
3--2 $S$(23) & 2.1181 & 48,690 & 22.2 & 0.25 & 0.12 & o \\
1--0 $S$(1)   & 2.1218 & 6,952 & 3.47 & 203.3 & 100.0 & \\ 
3--2 $S$(4)   & 2.1280 & 19,912 & 5.22 &  1.5 & 0.73 & \\
4--3 $S$(6)   & 2.1460 & 26,615 & 3.54 & 0.16 & 0.08 & \\
5--4 $S$(13) & 2.1528 & 39,539 & 5.08 & 0.20 & 0.10 & \\
2--1 $S$(2)   & 2.1542 & 13,151 & 5.60 & 8.7 & 4.2 &  \\
4--3 $S$(20) & 2.1630 & 46,783 & 22.8 & 0.08 & 0.04 & f,m \\
5--4 $S$(14) & 2.1634 & 40,948 & 7.53 & 0.08 & 0.04 & f,m \\
2--1 $S$(27) & 2.1790 & 52,643 & -  &  0.13 & 0.06 & \\
5--4 $S$(15) & 2.1818 & 42,379 & 10.4 & 0.23 & 0.11 & \\
4--3 $S$(5)   & 2.2010 & 25,624 & 3.22 & 1.1 & 0.51 & g \\
3--2 $S$(3)   & 2.2014 & 19,087 & 5.63 & 3.1 & 1.43 & g\\
7--5 $O$(7)   & 2.2047 & 36,590 & 4.87 & 0.13 & 0.06 & \\
8-6 $O$(5)   & 2.2107 & 39,221 & 8.81 & 0.08 & 0.04 & \\
4--3 $S$(21) & 2.2196 & 48,345 & 25.3 & 0.23 & 0.10 & \\
1--0 $S$(0)   & 2.2233 & 6,472 & 2.53 & 40.5 & 18.4 & \\
5--4 $S$(17) & 2.2433 & 45,275 & 16.1 & 0.27 & 0.12 & \\
2--1 $S$(1)   & 2.2477 & 12,551 & 4.98 & 16.2 & 7.2 & \\   
3--2 $S$(25) & 2.2663 & 51,939 & 30.7 & 0.14 & 0.06 & h \\
4--3 $S$(4)   & 2.2668 & 24,734 & 4.02 & 0.36 & 0.16 & h \\
3--2 $S$(2)   & 2.2870 & 18,387 & 5.63 & 1.6 & 0.69 & \\  
4--3 $S$(3)   & 2.3445 & 23,955 & 4.58 & 1.1 & 0.46 & m \\  
2--1 $S$(0)   & 2.3556 & 12,095 & 3.68 & 3.6 & 1.49 & s \\   
3--2 $S$(1)   & 2.3864 & 17,819 & 5.14 & 2.6 & 1.06 & m  \\
6--5 $S$(15) & 2.4030 & 45,526 & 12.1 & 0.17 & 0.07 & m \\ 
1--0 $Q$(1)  &  2.4066 & 6,149 & 4.29 & 127.5 & 51.1 &  s \\
1--0 $Q$(2)  &  2.4134 & 6,472 & 3.03 & 40.5 & 16.2 & s \\ 
1--0 $Q$(3)  &  2.4237 & 6,952 & 2.78 & 176.3 & 70.0 & s \\    
1--0 $Q$(4)  &  2.4375 & 7,585 & 2.65 &  51.5 &  20.3 & \\  
\hline
\end{tabular}
\end{center}
\scriptsize
$^a$ in 0\farcs6~$\times$~0\farcs9 (RA $\times$ Dec) region shown in  Fig. \ref{S1line_map}\\
$^b$ \citet{dab84}, \citet{wol98} \\
$^c$ \citet{tur77} \\
$^d$ Arb. units; 1$\sigma$$\approx$0.04 for weak lines. \\
$^e$ for $A_K$=1.09 mag;  scaled to 1--0 S(1) = 100; 1 unit is 1.0 $\times$ 10$^{-4}$ erg cm$^{-2}$~sr$^{-1}$\\
$^f$ blend, assume equal contributions \\
$^g$ blend, see section 4.2 \\
$^h$ blend, deconvolved \\
$^o$ significant contamination by OH sky emission line\\
$^m$ line peak within one res. element of 0.5$<$$\tau$$<$1 telluric absorption line\\
$^s$  line peak within one res. element of $\tau$$>$1 telluric absorption line\\
\end{table}
\normalsize

\begin{table}[http]
\caption{Undetected H$_2$ Emission Lines}
\label{undetectedlines}
\begin{center}
\scriptsize
\begin{tabular}{|c|c|c|c|c|c|}
\hline
Line & Rest $\lambda$  & Upper Level & A & Predicted & Notes \\
ID & vac. $\mu$m$^a$ & Energy (K)$^a$ & 10$^{-7}$s$^{-1}$ $^b$ & Rel. Intensity$^c$ & \\
\hline
7--5 $O$(5)   & 2.0220 & 35,615  & 8.43 & 0.1 & s \\
6--4 $O$(7)   & 2.0297 & 32,713 & 4.06 &  0.1 & o,m \\
2--1 $S$(25) & 2.0348 & 49,191 &  -  & 0.1 & o,m \\
5--4 $S$(5)   & 2.3555  & 30,064 & 1.82 & 0.1 & d \\
4--3 $S$(2)   & 2.4355 & 23,296 & 4.78 & 0.2 & s \\
\hline
\end{tabular}
\end{center}
\scriptsize
$^a$ \citet{dab84}, \citet{wol98} \\
$^b$ \citet{tur77} \\
$^c$ scaled to 1--0 $S$(1) = 100 \\
$^d$ blend with 2--1 $S$(0) \\
$^o$ significant contamination from OH sky emission line\\
$^m$ line peak within one res. element of 0.5$<$$\tau$$<$1 telluric absorption line\\
$^s$  line peak within one res. element of $\tau$$>$1 telluric absorption line\\
\end{table}
\normalsize

\section{Initial Analysis Steps}

In this section we restrict our analysis to the areas of brightest line emission and to regions where the contribution from the blue-shifted velocity component (see Fig.~\ref{velocity}) to the total line intensity is insignificant, so that the emission is likely to originate in one contiguous region rather than two or more widely separated regions. This restricted area is a  $\sim$1\farcs5-wide swath extending from the leading edge of the bow shock, near its tip, to the southwest and is the mapped region in  Fig. \ref{Q3S1} and some subsequent figures.

Much of our of our analysis and discussion is based on the relative intensities of H$_2$ lines and on maps of several line intensity ratios: 1--0 $Q$(3)/1--0 $S$(1), 2--1 $S$(1)/1--0 $S$(1), 3--2 $S$(3)/ 2--1 $S$(1), 3--2 $S$(1)/2--1 $S$(1), and 4--3 $S$(3)/3--2 $S$(1), covering a wide range of upper energy levels. Prior to determining intensity ratios we binned the original NIFS reduced data cube, whose elements are 0\farcs05~$\times$~0\farcs05, into a coarser cubes whose elements cover 0\farcs15~$\times$~0\farcs15 and 0\farcs30~$\times$~0\farcs30 regions. The latter undersamples the seeing but gives sufficiently high signal-to-noise ratios  to produce meaningful line intensity ratios involving the weakest  of the above lines. Line strengths were determined by numerical integration across each line profile, as there is no continuum. 

\subsection{Extinction Correction}

Extinction by dust attenuates the H$_2$ lines in the $K$-band spectra by wavelength dependent amounts. Understanding the physical processes occurring in the molecular shock depends on reliable values of the relative intensities of lines from different energy levels, and could be skewed if extinction, including its variation across the line-emitting region, is not taken into account. Uncertainty in the exact form of the extinction law results in some residual uncertainty in the line ratios after the correction. In our choice of line ratios to analyze we have reduced this residual by selecting pairs of lines that are close in wavelength. We apply the extinction correction as follows.

\begin{figure}[h!]
\begin{center}
\includegraphics[width=0.8\textwidth]{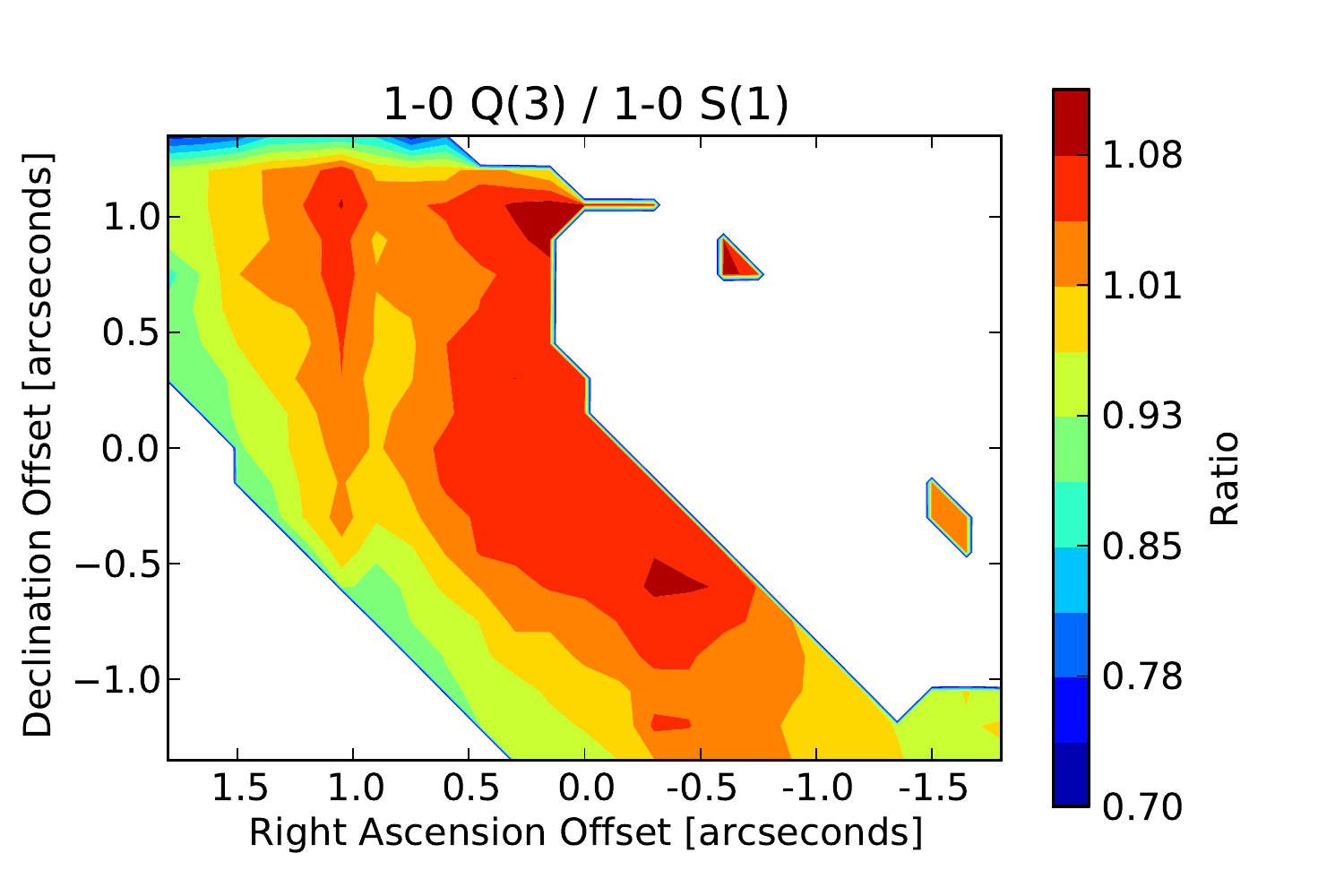}
\caption{The observed 1--0 $Q$(3)/1--0 $S$(1) line intensity ratio in the region of brightest line emission at and behind the shock front. Higher values of the ratio correspond to larger line of sight extinctions (see text).}
\label{Q3S1}
\end{center}
\end{figure}

We use the 1--0 $S$(1) and $Q$(3) emission lines of H$_2$, which have the same upper energy level and are widely separated in wavelength, to determine and correct for extinction.  (A second pair of lines, 1--0 $S$(2) and 1--0 $Q$(4), have a somewhat larger wavelength separation, but the former line is coincident with strong night sky OH emission and was judged unsatisfactory.) From the spontaneous decay rates of these lines  \citep{tur77,wol98}, the intrinsic $Q$(3)/$S$(1) line intensity ratio is 0.70.  The observed ratio ranges from 0.85 to 1.10, as can be seen in Fig. \ref{Q3S1}. We assume the extinction has the wavelength dependence $\lambda^{-1.7}$ \citep{ind05} and calculate its value in the $K$ band at 2.20~$\mu$m using equations from \citet{oco05},

\begin{equation}
A_K = \frac {\Delta} {(\frac{2.2}{\lambda_2})^{1.7} - (\frac{2.2}{\lambda_1})^{1.7}}
\end{equation}

\noindent where

\begin{equation}
\Delta = 2.512 log[\frac{F_{Q3}}{0.70F_{S1}}]
\end{equation}

\noindent with the F the observed line fluxes. Then the correction factor for a line at wavelength $\lambda$ is

\begin{equation}
f_{corr} = 2.512^{(A_{K}(\frac{2.2}{\lambda})^{1.7})}.
\end{equation}

We have scaled each line flux by the correction factor determined from the extinction measured at its location, using the above expression. In Table~\ref{lines} both the observed and extinction-corrected relative line strengths are presented; in the latter the strength of the 1--0 $S$(1) line is set to 100. The extinction in the $K$ band is found to vary from to 1.0 to 2.3 mag, corresponding to a range in visual extinction of 12-28 mag. In general the lowest extinctions are located near the leading edge of the shock.  This suggests that the leading edge of the shock, including its foreground flank as viewed from earth, is closer to breaking through the ambient cloud than the other regions of the shock observed here, and implies that the ambient cloud into which the outflow is penetrating is highly non-uniform.

\subsection{3--2 $S$(3) Blending Correction}

Of the H$_2$ lines that we employ for analysis of the excitation conditions in the shock, one, the 3--2 $S$(3) line at 2.201~$\mu$m, is blended with the 4--3 $S$(5) line; their wavelength difference corresponds to a velocity separation of only 55~km~s$^{-1}$.  We have used the 4--3 $S$(3) and 3--2 $S$(1) lines, which have almost the same energy difference as the blended lines, to estimate the relative contributions of 3--2 $S$(3) and 4--3 $S$(5) to the bended emission. For these two line pairs, using the upper level energies and spontaneous decay rates of the transitions and the level degeneracies $g_J$~=~ 21, 33, 45 for the upper rotational levels $J$~=~3, 5, 7, respectively, the ratios of the line strengths are
 
\begin{equation}
R_{\frac{4-3 S(5)} {3-2 S(3)}} = 0.78e^{-6537/T}
\end{equation}

\noindent and 

\begin{equation}
R_{\frac{4-3 S(3)} {3-2 S(1)}} = 1.425e^{-6137/T}
\end{equation}

\noindent so that 

\begin{equation}
R_{\frac{4-3 S(5)} {3-2 S(3)}} = 0.55e^{-400/T} R_{\frac{4-3 S(3)} {3-2 S(1)}}  \approx 0.50R_{\frac{4-3 S(3)} {3-2 S(1)}}
\end{equation}

\noindent for values of T of several thousand Kelvins or more. The observed value of R$_{4-3 S(3) / 3-2 S(1)}$  is typically 0.5, implying that 3--2 S(3) typically contributes 75\% of the flux of the blend at 2.201~$\mu$m. This is consistent with the relative intensities of these two lines in Table~\ref{lines}.

\begin{figure}[h!]
\begin{center}
\includegraphics[width=1.0\textwidth]{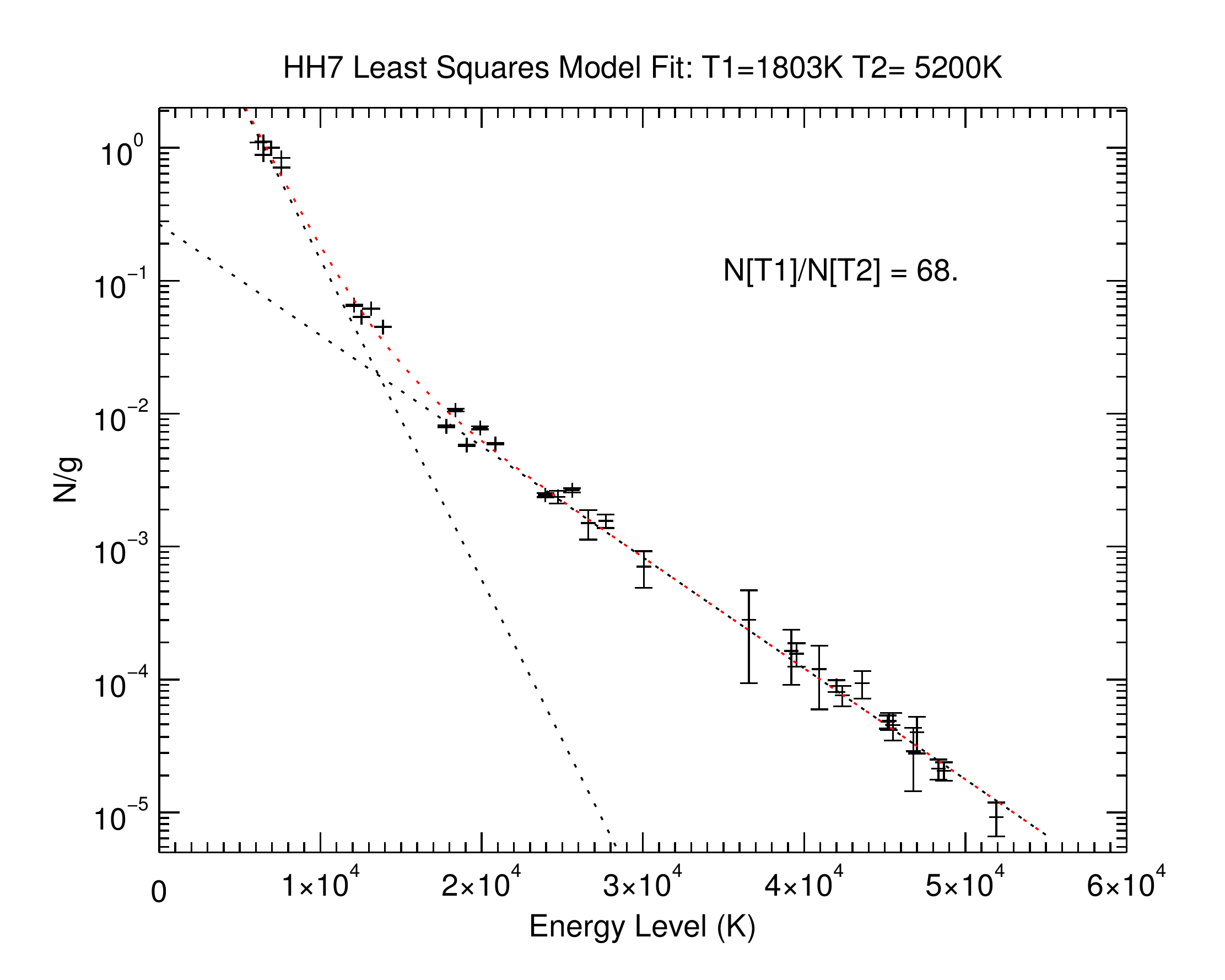}
\caption{Level column densities, divided by their degeneracies, $N_i/g_i$, plotted as a function of level energy, T$_i$, for the H$_2$ lines in HH7. The y axis, $N_i / g_i$,  is normalized to unity for the $(v,J) = (1,3)$ level (corresponding to the upper level of the 1--0 $S$(1) line). Error bars correspond to 1$\sigma$ values of 0.04 for the relative intensities listed in Table~\ref{lines}. The red curved dashed line, the sum of the two straight dashed lines, shows the best two-temperature LTE fit, for an ortho-to-para ratio of 3, as described in Section 5. The ratio of the column density of excited H$_2$ in the warm component (steeper straight line) to that of the hot component is 68.}
\label{colddenfit}
\end{center}
\end{figure}

\begin{figure}[h!]
\begin{center}
\includegraphics[width=0.56\textwidth]{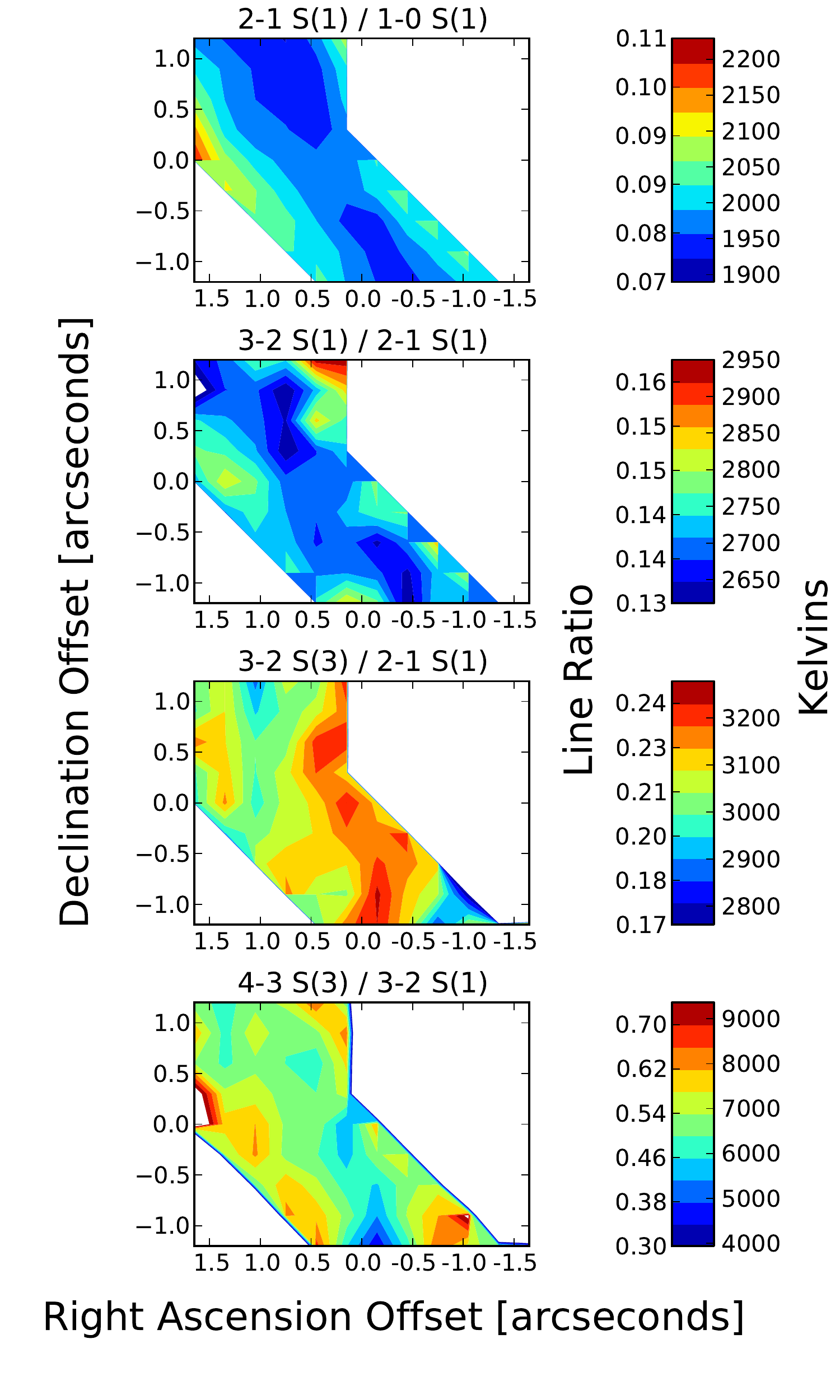}
\caption{Maps of line ratios and excitation temperatures in the region of brightest line emission, for  2--1 $S$(1)/1--0 $S$(1), 3--2 $S$(1)/2--1 $S$(1), 3--2 $S$(3)/2--1 $S$(1), 4--3 $S$(3)/3--2 $S$(1), after extinction correction. The colorbar indicates line ratios (left) and excitation temperatures (right).  See text for details.}
\label{ratiosmap}
\end{center}
\end{figure}

\begin{figure}[h!]
\begin{center}
\includegraphics[width=0.55\textwidth]{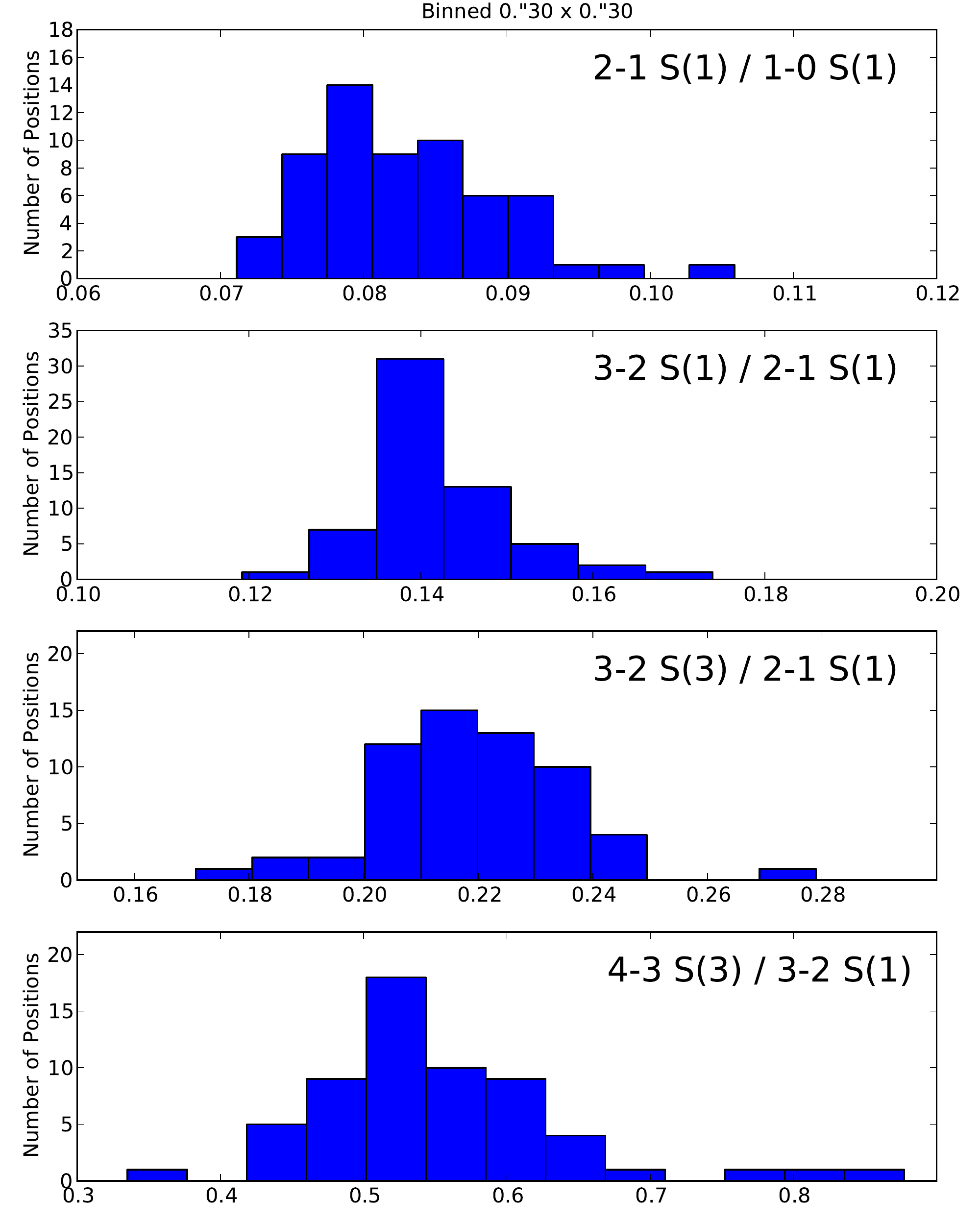}
\caption{Histograms of distributions of line intensity ratios 4--3 $S$(3)/3--2 $S$(1), 3--2 $S$(3)/2--1, $S$(1) 3--2 $S$(1)/2--1 $S$(1), 2--1 $S$(1)/1--0 $S$(1) in 0\farcs3~$\times$~0\farcs3 regions within the area of strong line emission shown in Fig. \ref{Q3S1}. The extinction correction has been applied.}
\label{ratioshist}
\end{center}
\end{figure}

\begin{figure}[h!]
\begin{center}
\includegraphics[width=0.65\textwidth]{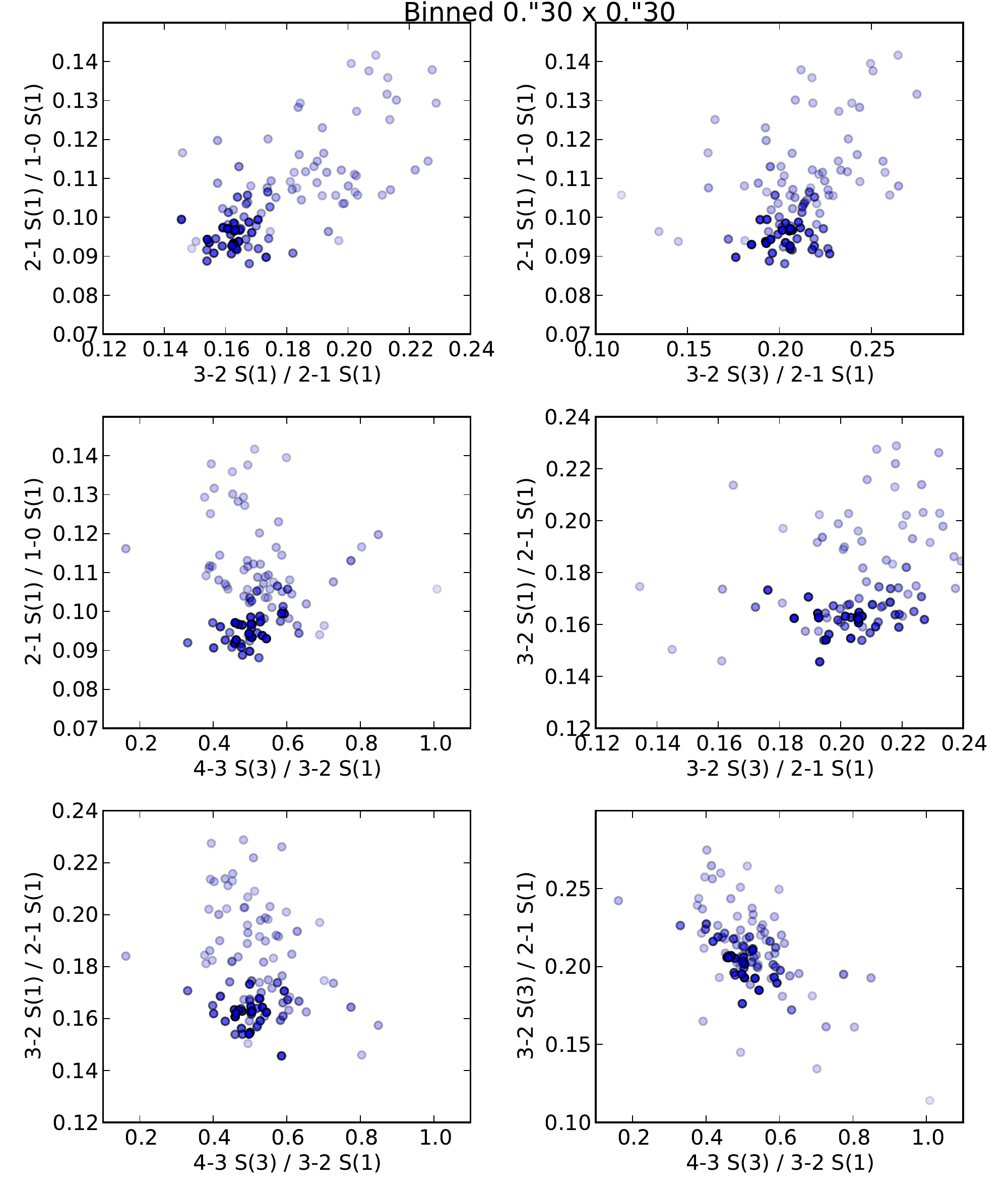}
\caption{Plots of pairs of line ratios shown in Fig. \ref{ratioshist}. The opacities of the data points scale with the strength of the 1--0 $S$(1) line; thus the lighter points have lower signal-to-noise ratios.}
\label{ratiosscat}
\end{center}
\end{figure}

\section{Level Populations}

\label{sec:analysis}

From the de-reddened relative intensities, $I_i$, listed in Table~\ref{lines}, level column densities, $N_i$, can be derived from
\begin{equation}
I_i = N_i A_i h \nu_i / 4  \pi,
\end{equation}

\noindent where $A_i$ is the radiative decay rate from the ro-vibrational level $i$ \citep[from][]{wol98}, and $\nu_i$ is the frequency of the transition \citep[from][]{tur77} .

A standard plot of level column density vs. energy is shown in Fig.~\ref{colddenfit}. Clearly, a single temperature Boltzmann distribution cannot account for all of the the observed line intensities, as the slope of the plot is significantly different at low and high energies.   Instead, a two-component Boltzmann distribution with temperatures $T_{warm}$ and $T_{hot}$ has been fit to the column densities, as follows:

\begin{equation}
N_i =  N_{i,warm} + N_{i,hot}
\label{eqn:fitformula}
\end{equation}
with
\begin{equation}
\frac{N_{i,hot}}{N_{hot}} = \frac{g_i e^{-T_i/T_{hot}}}{Z(T_{hot})},
\end{equation}

\noindent where $N_{hot}$ is the total column of hot gas, $g_i$ the level degeneracy, is 3$\times(2J+1)$ for ortho (odd $J$) states and 2$J$+1 for para (even $J$) states, and $T_i$ the energy level (in K).  The formula for  $N_{warm}$ is similar. $Z(T)$ is the partition function, given by
\begin{equation}
Z(T) = \frac{2T}{T_{rot}}(1 - e^{-T_{vib}/T})
\end{equation}

\noindent with rotational and vibrational constants $T_{rot} = 88$\,K and $T_{vib} = 6,000$\,K  \citep[see][]{bur89b}.

A  Levenberg--Marquard minimum-$\chi^2$ fit  to eqn.~\ref{eqn:fitformula} has been made using the \textit{lmfit} routine in IDL\footnote{http://www.exelisvis.com/ProductsServices/IDL.aspx}. Excellent agreement with the observations is obtained, as shown in Fig.~\ref{colddenfit}.\footnote{Note that for determining the minimum $\chi^2$--value we set $1 \sigma$ values for the line intensities to be 0.04, corresponding to the weakest lines identified in the spectrum in Table~\ref{lines}.} We find that $T_{warm}$ = 1,803~$\pm$~12~K and $T_{hot}$ = 5,200 $\pm$~12~K. To avoid being overly constrained by artificially high signal-to-noise ratios on the strong lines, since the dynamic range of the H$_2$ line strengths is so large, we set the maximum signal-to-noise ratio for any individual line to be 50 in the fitting analysis.  Of the line-emitting gas, 98.5\%\ is at 1,800~K and 1.5\%\ is at 5,200~K.  For individual lines the fraction of the emission arising from the hot and warm components depends on the upper level energy.  For instance, for the fiducial 1--0 $S$(1) line, 92\%\ of the line emission comes from the warm component and 8\% from the hot component. For the (unobserved) lower rotational levels of the ground vibrational state, nearly all of the H$_2$ is warm; for example 99.0\%\ of the H$_2$ in the $v$=0, $J$=3 state is in the warm gas. In contrast, for H$_2$ in levels of energy $\sim$40,000~K and higher less than 0.01\%\ is in the warm gas. The H$_2$ at energy levels near 13,650~K, which corresponds roughly to $v$=2, $J$=5 (the upper level of the 2--1 $S$(3) line at 2.0735~$\mu$m), is composed roughly equally of hot and warm gas.

\subsection{Comparisons with Previous Results}

As seen in Fig.~\ref{colddenfit} the accuracy and sufficiency of a two-temperature fit in describing the population diagram is not obvious unless lines arising from levels $T_i \geq 30,000$~K are observed. Such H$_2$ lines had not been detected previously in shocked regions. Similar behavior as seen here, exhibited in high velocity shocks elsewhere, could explain why the clear separation into two temperature components has not been made before; e.g., by \citet{bra88} and \citet{bur97}, who measured H$_2$ lines up to 25,000~K above the ground state in the brightest H$_2$ emission source in the sky, Peak 1 in the Orion Molecular Cloud.  Although those authors found an excess in level populations for the highest excitation lines they measured, above that expected for 2,000~K gas, they fit the observed level populations to those produced in planar J-shocks with cooling of the hot gas being described by a power law model that approximated that given by H$_2$ itself, rather than to cooling contributions from other species such as H$_2$O and [O I]. 

In the case of HH7 the two-temperature fit to the population diagram is consistent with the measurements of \citet{fer95}, who found the H$_2$ level populations could not be fit by a gas in LTE at a single temperature, and are also compatible with the first measurements made of the H$_2$ spectrum of this source by \citet{bur89a}.  The latter authors observed  the 1--0 and  2--1 $S$(1) lines in the $K$--band, as well as the 0--0 $S$(13) and 1--0 $S$(7) lines at 3.8$\mu$m.  From the former they estimated an excitation temperature $T_{ex} \sim 1,900^{+400}_{-200}$\,K, assuming a differential extinction between the two $S$(1) lines of $\Delta \tau = 0.1$, equivalent to  $A_K \sim 1.3$\,mags.  From the 0--0 $S$(13) and 1--0 $O$(7) lines, which are well separated in energy level (17,450 vs.\ 8,300\,K) but are at almost the same wavelengths and thus suffer identical extinctions, they determined $T_{ex}~\sim$ 3,100$\pm300$\,K.  The 0--0 $S$(13) line does not arise from a sufficiently high energy level for the hottest gas to be clearly identified, but nevertheless the presence of that gas is indicated by the higher excitation temperature.

\subsection{Missing and Unidentified Lines}

We have applied the two-temperature fit to predict the strengths of all H$_2$ lines, making use of the energy levels and transition probabilities from \citet{tur77} and \citet{wol98}, respectively. Five undetected lines, shown in Table~\ref{undetectedlines}, are predicted to have intensities $\sim 0.1$ units (i.e.\ $\sim 2\sigma$ in the flux scale of Table~\ref{lines}).  Of these lines, 2--1 $S$(25) is blended with the strong 1--0 $S$(2) line, and 5--4 $S$(5) is blended with the strong 2--1 $S$(0) line, while  the 7--5 $O$(5), 6--4 $O$(7) and 4--3 $S$(2) lines fall in regions of the spectrum badly affected by telluric absorption. Thus it is not expected that any of these lines would have been detected in our data. 

Only one emission line in the spectrum is unidentified. It is at 2.1044~$\mu$m and is only marginally detected.  The closest H$_2$ lines to this wavelength are 9--7 $Q$(3) at 2.1006~$\mu$m and 7--5 $O$(6) at 2.1090~$\mu$m, each of which are $\sim$10 spectral resolution elements distant and neither of which is predicted by our model to be detectable.  Thus this unidentified line, if real, is not due to H$_2$.

\section{Excitation Temperatures}

The excitation temperature, T$_{21}$, of a pair of H$_2$ lines is determined from their extinction-corrected intensities, $I$, upper level energies, $E$, spontaneous decay rates, $A$, and the degeneracies of the rotational levels, $g$, by

\begin{equation}
\frac{I_2}{I_1} =
        {\frac {A_2g_2\lambda_1e^{\frac{-E_2}{T_{21}}}}{A_1g_1\lambda_2 e^{\frac{-E_1}{T_{21}}}}}~.
\end{equation}

\noindent Maps of the excitation temperature for selected H$_2$ line pairs (2--1 $S$(1)/1--0 $S$(1), 3--2 $S$(1)/2--1 $S$(1), 3--2 $S$(3) / 2--1 $S$(1), and 4--3 $S$(3) / 3--2 $S$(1)), based on values in 0\farcs3~$\times$~0\farcs3 regions, are shown in Figure \ref{ratiosmap}, and histograms of each line ratio are displayed in Fig.~\ref{ratioshist}. Scatter plots of pairs of line ratios are shown in Fig.~\ref{ratiosscat}. In the latter figure the opacity of a data point scales with the strength of the 1--0~$S$(1) line, and thus roughly with the strengths of all of the lines. Although each ratio apparently has a range of $\pm$ 25\%--50\%  the values for the stronger lines cluster within much smaller ranges, centered on approximately 0.10 for 2--1 $S$(1)/1--0 $S$(1), 0.16 for 3--2 $S$(1)/2--1 $S$(1), 0.20 for 3--2 $S$(3)/2--1 $S$(1), and about 0.45 for 4--3 $S$(3)/3--2 $S$(1). Because the strengths of all of these lines scale roughly with the strength of the 1--0 $S$(1) line, and the signal-to-noise ratios on the 3--2 and 4--3 lines are not very high, we believe that much of the scatter is due to noise (random fluctuations and/or cosmic rays that produced slightly spurious data points not obvious enough to be removed or corrected). Our derived 2--1 $S$(1)/1--0 $S$(1) flux ratios in the bright portion of the shocked gas are somewhat lower that found by \citet{smi03}.

The above values of the line ratios correspond to excitation temperatures of 2150~K, 2920~K, 2960~K, and 5300~K, respectively. The trend, that ratios involving transitions from higher energy levels are associated with higher excitation temperatures, is similar to that seen previously in HH7 and elsewhere, as discussed in Section 5.1. However, Fig.~\ref{colddenfit} makes it clear in that in HH7 the excited H$_2$ producing the $K$-band lines is not distributed over a wide range of temperatures but instead is predominantly at only two temperatures, $\sim$2000~K and $\sim$5000~K, and thus that the different excitation temperatures reflect the different contributions to the line emission from the two components.

\section{Discussion}

\subsection{Introduction}

The most important result from these observations is the discovery of a significant contribution to the H$_2$ line emission from molecular gas in LTE at about 5,000~K, a much higher temperature than has been previously found in shocked molecular gas. The vast majority of the $K$-band line emission from HH7 comes from gas at $\sim$2000~K, which can be accounted for by a continuous-shock \citep{smi03} in which a magnetic precursor associated with the high speed outflow (from SSV~13) gradually accelerates the ambient molecular gas, with the result that the collisions between the bulk of the outflowing ambient gas does not dissociate the molecules. However, the origin of the much smaller amount of emitting H$_2$ at $\sim$5,000~K  is less clear.

\subsection{Possible origins of the 5,000~K H$_2$}

\subsubsection{Fluorescence}

Previously \cite{fer95} suggested the presence of a fluorescent component to the H$_2$ line emission in HH7, based on measured fluxes of a few lines from $v$=1--4 and $J<5$ upper levels with energies up to $\sim$25,000~K. However, our analysis of the spectrum of HH7 shows that the emission is clearly thermal in nature, with the column densities showing no significant deviation from the fit to a  two-temperature component Boltzmann distribution, up to the dissociation energy of the H$_2$ molecule at 4.5\,eV\@.  A significant contribution from UV-excited H$_2$ would result in significantly higher vibrational than rotational temperatures \citep[e.g.][]{bla87} derived from the level column densities and would have created a much more scattered distribution of data points in Fig.~\ref{colddenfit}. In dense gas collisional re-distribution can lead to a thermalization of lower vibration levels, giving rise to an apparent thermal distribution in v=0, 1, and 2 with $T_{ex} \sim2000$\,K, but the higher vibrational levels, which spontaneously decay more rapidly, would still retain the signature of the non-thermal fluorescent cascade. We note that the model of HH7 by \citet{smi03}, based on observations of the 1-0 $S$(1) and 2--1 $S$(1) lines, does not predict significant fluorescent line emission. 

\subsubsection{Heating by ionized gas}

Sufficiently fast shocks ($> $50 km s$^{-1}$) are expected to result not only in nearly complete dissociation of H$_2$ \citep{che82}, but also ionization of atomic hydrogen. This would result in recombination line emission from atomic hydrogen. Any surviving H$_2$ in the this region could be heated to temperatures of many thousands of Kelvins. However, no hint of Br~$\gamma$ line emission at 2.166~$\mu$m is present in Fig.~\ref{binned_spec}; its intensity must be at least 1,000 times less than that of the 1-0 $S$(1) line. Our coarse examination of the data cube shows little or no evidence for H~I Br~$\gamma$ line emission elsewhere, although at one or two locations within the bow shock  Br~$\gamma$ may be present, at flux levels five times lower than the adjacent H$_2$ 2--1 $S$(2) line and 100 times weaker than the fiducial 1--0 $S$(1) line at those locations. No Br~$\gamma$ line emission is present along the western ridge of emission at high negative velocities, which we have identified with the shock in the jet, where the shock conditions may be more extreme than elsewhere. 

In an analysis of observations of the molecular shock in the supernova remnant IC 443 \citet{bur90} obtained a lower limit of 40 on the H$_2$ 1-0 $S$(1) to Br~$\gamma$ line intensity ratio.  That limit allowed them to rule out fast J-shocks ($v$~$>$~100~km~s$^{-1}$) in the molecular gas, as these would have produced detectable levels of Br~$\gamma$ line emission.  However, neither slower J-shocks, nor C-shocks, could be ruled out by these authors in IC 443.  Our lower limit to the above ratio in HH7 is considerably higher. We similarly conclude that fast dissociative J-shocks are not present in HH7.

\subsubsection{Formation pumping}

Formation pumping is another possible means of contributing high excitation lines to the H$_2$ emission spectrum.  After being formed on dust grain surfaces \citep{hol71}, H$_2$ molecules are ejected in excited states following the release of their 4.5\,eV binding energy.  Some of the energy must overcome the grain surface potential, and the rest is distributed between kinetic energy and internal energy (in ro-vibrational levels), with the latter being emitted largely in lines as the molecules relax, as first modeled by \citet{bla87}.  

The formation process and the properties of the ejected H$_2$ depend on the nature of the grain surface. \citet{dul86} considered the surfaces to be highly defected silicates to which the molecule is moderately strongly bound. A later study \citep{dul93} considered H$_2$ formation on amorphous H$_2$O ice (where it is weakly bound) and on aromatic carbon molecules (e.g.\ polycyclic aromatic hydrocarbons (PAHs) or hydrogenated amorphous carbon (HACs), where it is strongly bound). In the former, which presumably does not apply to recently shock-heated matter, H$_2$ would be released from the surface in highly excited states, with the binding energy virtually all going into internal energy. In the latter case, which may also not apply, most of the binding energy would be distributed through the many degrees of freedom of the aromatic molecule, leaving the newly-formed H$_2$ in a low-excitation state. Subsequent investigators \citep[e.g.,][]{leb95, dra96,tak01} have also studied formation on these surfaces.  

Currently there is no clear view as to the level distribution which the newly formed molecules would have. \citet{dul86} predicted that the H$_2$ molecules are ejected into the $v = 6$ level, containing about 1.5\,eV of the bond energy. Observations in the Northern Bar of the M17 photodissociation region (PDR) by \citet{bur02} were interpreted by the authors as the signature of H$_2$ formation pumping, through their detection of emission in the 1.7326~$\mu$m 6--4 $O$(3) line, and based on the clear difference of its emission morphology from that of lower-excitation fluorescent H$_2$ lines in the region.  That line originates from a level 31,055\,K ($\sim$2.7~eV) above ground state.  However considerably different level distributions have been proposed by the other investigators cited above. 

We have found that the highly excited H$_2$ is roughly in LTE at $\sim$5,000~K. Although it is possible that H$_2$ molecules are released from the grains with a thermal internal energy distribution, we consider it more likely they are ejected in a highly excited, but non-thermally distributed ro-vibrational levels, as in many of the theoretical calculations.  If so they would need to be quickly collisionally re-distributed in the post-shock gas, to produce the observed  thermal population distribution at $\sim$5,000~K.  That temperature is equivalent to 0.45\,eV and would represent about one-third of the internal energy if the H$_2$ molecules are released with the 1.5\,eV in the \citet{dul86} model, and about 15\% of the energy if all H$_2$ were released into the 6--4 $O$(3) level seen in the M17 PDR\@.  Alternatively, if the ejected molecules do have energies of several eV they might be able to collisionally heat a larger mass of gas, for instance $\sim 6$ times more if ejection occurred with H$_2$ in $v=6$ levels.  This redistribution must also happen rapidly, since the cooling time for highly excited levels is short; the radiative decay rates for high energy states are relatively large. For instance the 6--4 $O$(3) line has a decay rate four times larger than the 1--0 $S$(1) line, and a corresponding cooling timescale of one week. If the density in the post-shock gas is at least 10$^5$~cm$^{-3}$ and is at temperatures of $\sim$5,000~K rapid collisional cooling of the hot, dense newly formed H$_2$ will occur on timescales of at most a few days.

\section{Future Observations and Calculations}

These results, particularly those relating to the existence of the hot H$_2$, pose many questions requiring additional observations and calculations. Foremost among them are determining the uniqueness of $\sim$5,000~K gas and understanding its origin. First, does the temperature of this hot gas and its abundance relative to the warm gas vary with location in the HH7 shock? As pointed out, the weak lines are difficult to observe far from the brightest portions of the shock. However, we have only observed a part of the brightest region. In particular, it could be revealing to observe the $K$-band H$_2$ spectrum in the cap of the bow shock, where one would expect that there is more dissociation. If the hot H$_2$ is newly reformed it might constitute a higher faction of the line emitting gas there. Second, are the temperature of the hot gas, and indeed its existence unique to HH7?  We expect that the answer to the second part of this question is no, as lines from energy levels as high as 25,000~K have been seen elsewhere in collisionally excited H$_2$, as discussed earlier, and reveal a similar dependence of excitation temperature on upper energy level as found in HH7. We have very recently obtained deep $K$ band spectra of shocked H$_2$ in the OMC-1 outflow extending from near its Peak~1 through one of the bright H$_2$ ``bullets" \citep{all93,ted99}, which should shed light on both parts of this question. Third, could spectra at higher velocity resolution than presented here of lines from a wide range of upper level energies provide clues to the relationship of the hot component to the warm component? 

On the theoretical side, if the hot H$_2$ is recently reformed on grains it is important to determine its formation rate and its initial energy level distribution as it leaves the grains, and to understand the competition between energy loss via line emission and via collisions. If the internal energy is redistributed from a higher temperature distribution or from an initial non-LTE initial distribution the resulting (observed) temperature is likely to be sensitive to the physical conditions in the shocked gas, and in particular its density.  Finally there is even the question of how a single hot temperature would result from any re-distribution. The high quality of the two-temperature fit suggests that, if the hot H$_2$ in HH7 is reformed H$_2$, it must cool quickly from $\sim$5,000~K to $\sim$2,000 K so that intermediate temperatures are not detected in a population diagram. If that is not the case then the hot H$_2$ would seem to have to be located in a separate environment that is physically detached from the bulk of the post-shock gas.

\bigskip
\bigskip

This research is based on observations obtained at the Gemini Observatory, which is operated by the Association of Universities for Research in Astronomy, Inc., under a cooperative agreement with the NSF on behalf of the Gemini partnership: the National Science Foundation (United States), the National Research Council (Canada), CONICYT (Chile), the Australian Research Council (Australia), Minist\'{e}rio da Ci\^{e}ncia, Tecnologia e Inova\c{c}\~{a}o (Brazil) and Ministerio de Ciencia, Tecnolog\'{i}a e Innovaci\'{o}n Productiva (Argentina). We thank the staff of Gemini Observatory for its support of this research and Richard McDermid for assistance with the data reduction.  We also thank Peter Brand for helpful discussions regarding this research, as well as for all that he has taught us and that we have learned together with him over several decades.

\end{document}